# Ranking Spaces for Predicting Human Movement in an Urban Environment

Bin Jiang

Department of Land Surveying and Geo-informatics
The Hong Kong Polytechnic University, Hung Hom, Kowloon, Hong Kong
Email: bin.jiang@polyu.edu.hk

**Abstract**
A city can be topologically represented as a connectivity graph, consisting of nodes representing individual spaces and links if the corresponding spaces are intersected. It turns out in the space syntax literature that some defined topological metrics can capture human movement rates in individual spaces. In other words, the topological metrics are significantly correlated to human movement rates, and individual spaces can be ranked by the metrics for predicting human movement. However, this correlation has never been well justified. In this paper, we study the same issue by applying the weighted PageRank algorithm to the connectivity graph or space-space topology for ranking the individual spaces, and find surprisingly that (1) the PageRank scores are better correlated to human movement rates than the space syntax metrics, and (2) the underlying space-space topology demonstrates small world and scale free properties. The findings provide a novel justification as to why space syntax, or topological analysis in general, can be used to predict human movement. We further conjecture that this kind of analysis is no more than predicting a drunkard's walking on a small world and scale free network.

**1. Introduction**

A space, in the context of this paper, refers to a small scale space that is small enough to be perceived from a single vantage point. From this point of view, a city or an urban environment being part of a city is too big to be viewed from a single vantage point, and it is constituted by many (small scale) spaces. This way a city can be abstracted as a connectivity graph or space-space topology in which nodes and links represent individual spaces and their intersections respectively. Similarly, a building complex can be abstracted as a space-space topology in which nodes represent individual rooms or corridors, and links represent the doors. The space-space topology formed at either a city or a building level is an interconnected whole in which each space has a different topological status. For instance, some of the spaces are well connected (with a high degree of connectivity), while others are less so. This is the fundamental principle of space syntax (Hillier and Hanson 1984, Hillier 1996), which consists of topological representations forming a space-space topology on the one hand and some defined space syntax metrics on the other. A major finding of space syntax has been the prediction of human movement using some space syntax metrics.

How one space is connected to other spaces in the space-space topology illustrates its status in the whole at a local level. Assuming a frog randomly jumping from one node to its neighboring nodes (or successors) in the connectivity graph, the chance that the frog will visit individual nodes appears to be decided by the connectivity of the nodes. This is simply based on the theory of Markov chains (e.g., Ching and Ng 2006). Although human movement is not as simple as the frog's jumping, intuitively the connectivity of the nodes does show some level of correlation to pedestrian and vehicle flows. However, researchers in the space syntax community claimed, through enormous empirical studies (Hillier et al. 1993, Penn et al. 1998), that human movement both pedestrian and vehicle can be predicted by local integration, a node's path length within two steps, rather than the connectivity. As shown in the following section 2.1, local integration is a normalized closeness centrality initially developed in social network analysis (Scott 2000).

Local integration is the default indicator for human movement, but why it can be so seems unclear. More specifically why the integration or closeness centrality can be used to predict human movement is not well explained and justified. This has been a continuous debate in the space syntax community as to why space syntax works. In response to this, Hillier (1999) argued that people tend to take paths that minimize trip length or maximize trip efficiency. However, it is still not justified as to why closeness centrality at a local level instead of a global level if to maximize trip efficiency. Turner (2007) has recently challenged the issue, and proved that betweenness centrality is a better metric for predicting human movement. Turner's work has further argued that the conventional axial representation may be better replaced by centre-line street network.

In this paper, we apply an extended PageRank algorithm, weighted PageRank (Xing and Ghorbani 2004), to the space-space topology for ranking the individual spaces, and find surprisingly that the PageRank scores are significantly correlated to human movement both pedestrian and vehicle in four observed areas of London. The correlation is even slightly better than that of local integration. This finding provides a novel justification as to why space syntax or topological analysis can be used to predict human movement at a certain confidence level. Given another illustrated fact that the underlying topology exhibits small world and scale free properties, we conjecture that this kind of analysis is no more than predicting a drunkard's walking on a small world and scale free network. A word about PageRank. PageRank algorithm is the key technology behind the phenomenal success of Google search engine (Page and Brin 1998). It has been proved to be efficient and effective in ranking pages of the World Wide Web. PageRank has also been used for ranking journal status (Bollen et al. 2006) and the importance of publications (Chen et al. 2006). For a more comprehensive overview of PageRank and its applications, the reader may refer to the research monograph by Langville and Meyer (2006).

The remainder of this paper is organized as follows. Section two introduces several metrics for ranking nodes of a graph, including space syntax metrics, small world and PageRank metrics. Section three presents our experiments with the aims of examining (1) small world and scale free properties of the London axial map and four areas around central London, and (2) the correlation between the observed pedestrian and vehicle flows and the weighted PageRank scores, in comparison to the space syntax metrics. Based on the experiments, section four provides a justification of our findings, i.e., a never-get-stuck-or-tired random walker who has a preference while jumping to its neighboring nodes or successors. Finally section five concludes the paper and points to our future work.

## 2. Topological metrics for ranking spaces

Before introducing the metrics for ranking spaces, let us take a look at how an urban environment (Figure 1a) can be topologically transformed into a connectivity graph or space-space topology based on the space syntax principle. As briefly mentioned early, an urban environment, in particular the part of the environment in which people can freely move around, can be partitioned into many small scale spaces. The spaces can be approximated by individual axial lines (Figure 1b). The transformation from the axial map to the connectivity graph (Figure 1c) seems a straightforward operation, i.e., the axial lines and line intersections being respectively the nodes and links of the graph. For the sake of simplicity, we adopt the fictive urban system for illustrating the basic principle of space syntax. In practical exercises, the generation of an axial map for a reasonably large city is a very tedious process, as there is no efficient automatic solution yet. The limitation of space syntax in this kind of representation has been well argued in the space syntax community (e.g. Ratti 2004, Batty 2004, Jiang and Claramunt 2002), and alternative representations have been put forward (e.g., Jiang and Claramunt 2002, Jiang and Claramunt 2004) to overcome the shortcoming. Recent work by Jiang and Liu (2008) has indeed proved that street-based representations are superior to the axial map in predicting traffic flow. However, this paper does not deal with the representation issue; instead we concentrate on metrics by assuming an underlying topology is already correctly generated.

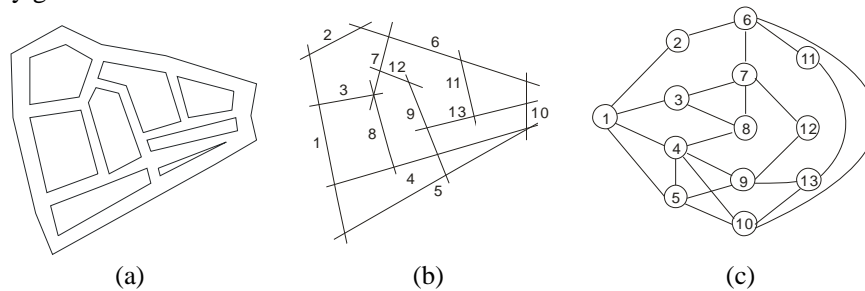

(a)  (b)  (c)

Figure 1: A fictive urban system (a), its axial map (b) and connectivity graph (c)

## 2.1 Space syntax metrics

Connectivity is the simplest metric for assessing ranking of nodes within a connectivity graph. It equals the number of directly linked or neighboring nodes. A second metric of space syntax is called integration, which is based on the concept of depth. For any particular node in the connectivity graph, the shortest distance (or steps) away from the node is denoted by $s$ ($s$ is an integer), the number of nodes with the shortest distance $s$ is denoted by $N_s$, and the maximum shortest distance is denoted by $k_i$. The $k_i$ varies from one node to another, but the longest shortest distance is the diameter of a graph. Using the expression $\sum_{s=1}^{k} s \times N_s$, both connectivity and depth are defined as follows:

$$\sum_{s=1}^{k} s \times N_s = \begin{cases} connectivity & iff \ s = 1 \\ local \ depth & iff \ 2 \leq s \leq k_i \\ global \ depth & iff \ s = k \end{cases} \quad [1]$$

$s=2$ implies that we consider those lines that are within two steps from an axial line. We can remark that *connectivity* is equivalent to *local depth* when $s = 1$. Taking the above node 1 as an example, this node has four directly linked nodes, so the connectivity is 4. Apparently it is a local property. In contrast, a global depth assesses how far it is from a node to all other nodes, at a global scale so to speak. For example, node 1 has four nodes in step one, five nodes in step two, and three nodes in step 3, thus the sum $4 \times 1 + 5 \times 2 + 3 \times 3$ indicates the global depth. Instead of all steps, if we consider those nodes that are within two steps, the sum $4 \times 1 + 5 \times 2$ gives the local depth.

The depth either local or global divided by the number of nodes involved minus one leads to local or global mean depth (MD). The integration used in space syntax is a metric for measuring relational asymmetry, either by *Relative Asymmetry* (RA) or by *Real Relative Asymmetry* (RRA). They are given as follows:

$$RA(i) = \frac{2(MD-1)}{(n-2)} \quad [2]$$

and,

$$RRA(i) = \frac{RA(i)}{D_n} \quad [3]$$

where $n$ is the number of nodes involved, and $D_n$ is given by

$$D_n = \frac{2\left\{ n\left[ \log_2\left(\frac{n+2}{3}\right) - 1 \right] + 1 \right\}}{(n-1)(n-2)}. \quad [4]$$

Note that for global integration, $n$ is the total number of nodes within the connectivity graph, while for local integration $n$ is the number of nodes within two steps from a particular node. In space syntax software like Axman[1] and Axwoman[2], integration is the reciprocal value of RRA. This means that the bigger the integration value the more integrated the axial line is. Integration is a de fact closeness centrality, if we forget the normalization process. The reader is encouraged to consult the space syntax literature (e.g., Park 2005) for more details about the integration definition.

---

[1] a space syntax program based on Macintosh computer, developed by Bartlett school of University College London.

[2] a suite of freeware space syntax programs implemented as an extension of ESRI GIS family or as a standalone application and Java applet; for downloading refer to http://www.hig.se/~bjg/Axwoman.htm

**2.2 Small world metrics**
The above concept of global depth has another name, called path length, which is one of the two small world metrics. It represents a degree as to how a node separates from all other nodes. The average path length of all individual nodes is that of the connectivity graph, indicating a separation between two randomly chosen nodes. For example, the above node 1 has a path length of $4 \times 1 + 5 \times 2 + 3 \times 3$ (which reads as 4 nodes in 1 step, 5 nodes in 2 steps, and 3 nodes in 3 steps), indicating an extent as to how it is far from all other nodes. A node with a smaller path length tends to be more centralized, as it is closer to every other node. For most real world networks, their path lengths are very short, approaching the path length of their equivalent random graph. It is often used to verify whether or not a graph has a small separation.

Another small world metric is the clustering coefficient, proposed by Watts and Strogatz (1998). It is defined by the probability that two neighbours of a given node are linked together, and measured by a ratio of actual links to all possible links, among the neighbours of the particular node.

$$C(i) = \frac{\text{\# of actual edges}}{\text{\# of possible edges}} \quad [5]$$

Again taking the above node 1 as an example; among its four neighbours, actual links are 1 (between nodes 4 and 5), and possible links are 6, so the clustering coefficient for the node is 1/6. The clustering coefficient can be used for ranking individual nodes in terms of local efficiency; the higher the clustering coefficient of a node, the more efficiently a message is diffused among the neighbours of the particular node. In the context of the paper, the two metrics of individual nodes are summed up to be the metrics of relevant graphs for verifying a small world structure.

**2.3 PageRank metrics**
PageRank is a key technology behind the Google search engine that decides the relevance and importance of individual web pages. Its computation is through a web graph in which nodes and links represent individual web pages and hotlinks. Obviously, the web graph is a directed graph, i.e., a hotlink from page A to B does not imply another hotlink from B to A. The basic idea of PageRank is that a highly ranked node is one that highly ranked nodes point to (Page and Brin 1998), apparently a recursive definition. PageRank is used to rank individual web pages in a hyperlinked database. It is defined formally as follows:

$$PR(i) = \frac{1-d}{n} + d \sum_{j \in ON(i)} \frac{PR(j)}{n_j} \quad [6]$$

where $n$ is the total number of nodes; $ON(i)$ is the outlink neighbors (i.e., those nodes that point to node $i$); $PR(i)$ and $PR(j)$ are rank scores of node $i$ and $j$, respectively; $n_j$ denotes the number of outlink nodes of node j; $d$ is a damping factor, which is usually set to 0.85 for ranking web pages.

The PageRank of a node relies on that of nodes that point to, i.e., a highly ranked node is one that highly ranked nodes point to. This is the beauty of the PageRank algorithm. The computation of PageRank is carried out iteratively starting with evenly distributed scores for individual nodes and ending when a convergence is reached. The true meaning of the scores is their relative value, rather than an absolute value, for the ranking purpose. PageRank is nicely justified as a random surfer who continuously with probability $d$ clicks hotlinks from one page to another. The random surfer jumps with probability *(1-d)* from one page to a randomly chosen page, or jumps with probability one to a randomly chosen page in case there is no hotlink for a page. PageRank is the stationary probability of a Markov chain (*cf.* Langville and Meyer 2006), and it is a variant of eigenvector centrality (Bonacich 1987) defined in social networks.

The above definition of PageRank has one problem, i.e. the PageRank of a node at any iteration is evenly divided over the nodes to which it links (or outlink nodes). For example, the PageRank of node 1 is supposed to propagate into its four linked nodes 2, 3, 4 and 5. According to equation [6], each one of the four nodes gets the same amount of PageRank contributed by node 1. However, the propagation of the PageRank should follow an uneven rule, i.e., the more popular nodes tend to get a higher proportion. In other words, it is proportional to the inlink connectivity[3] (or popularity) of the linked nodes. This is exactly the basic motivation of weighted PageRank (Xing and Ghorbani 2004).

The weighted PageRank is defined as follows:

$$PR(i) = \frac{1-d}{n} + d \sum_{j \in ON(i)} PR(j)W(j) \qquad [7]$$

where weight $W(j)$ is added to propagate a PageRank score from one particular node $i$ to its outlink nodes. This is different from equation [6], where a PageRank score is evenly divided among its outlink nodes.

The weight $W(j)$ represents relative popularity of node $j$ among its counterparts, and it is defined as follows:

$$W(j) = \frac{w(j)}{\sum w(k)} \qquad [8]$$

where $k$ is counterpart nodes of $j$, $w$ is the weight for individual links, indicating their relative popularity based on the percentage of inlinks and outlinks.

To illustrate the ideas behind PageRank and weighted PageRank metrics, we take node 1 for example by assuming the graph illustrated by Figure 1c is a web graph, and a web surfer is currently at page 1 (or equivalently node 1). He or she would click equally among the 4 linked pages, according to the standard PageRank definition. In other words, the web surfer follows page 1's successor in the same probability of 25%. However, according to the weighted PageRank, he or she would give a high priority to node 4, followed by nodes 5, 3 and 2. The corresponding probabilities would be 5/14, 4/14, 3/14 and 2/14. With the weighted PageRank definition, the web surfer has some priority agenda. This is also the key difference between the two PageRank metrics. It should be noted that for an undirected graph, the standard PageRank is identical to its connectivity in ranking. Thus it makes little sense to compute the standard PageRank for undirected graphs.

Initially PageRank is defined for web graphs that contain dangling nodes, representing the web pages that have inlinks but not outlinks (e.g., image files). However when it is applied to ranking spaces for an urban environment, the connectivity graph or space-space topology is a connected and undirected graph, thus no dangling nodes involved. In the following section, we will report our experiments to illustrate that (1) urban street networks demonstrate small world and scale free structures; (2) human movement is significantly correlated to the weighted PageRank, slightly better than to the local integration. These two illustrated facts together will lead to a novel justification as to why human movement can be predicted by topological metrics.

## 3. Experiments with three central London areas

### 3.1 The datasets
The datasets used in our experiments were studied in previous work on modeling human movement (Hillier et al. 1993, Penn et al. 1998). The London axial map forms an underlying topology, which was generated by taking each individual street or space as an axial line. The kind of axial map used is supposed to have a least number of longest axial lines. The London axial map consists of 17321 lines, each of which represents approximately a space in reality (Figure 2a). The axial map provides a skeleton view of the world city, and it can be represented and analyzed topologically. Three local areas in central London, namely Barnsbury, Clerkenwell, and South Kensington (Figure 2b), were selected for a more detailed investigation, in comparison to the entire London. Another dataset about human movement both pedestrian and vehicle in central London was obtained for verifying correlations between ranking metrics and human movement. These datasets are publicly available at http://eprints.ucl.ac.uk/archive/00001398/. A major reason why we adopt the datasets is that the datasets have been well studied previously (e.g. Hillier and Iida 2005, Carvalho and Penn 2004), becoming benchmark datasets in the space syntax community. Our aim is to investigate whether or not any alternative metric such as PageRank can be a better indicator for human movement in an urban environment, in comparison to the space syntax metrics. Furthermore, we intend to provide a novel justification as to why space syntax or topological analysis can be used to predict human movement. Our justification is based on the two factors: (1) the underlying topology that demonstrates small world and scale free properties, and (2) a random walker who gives a higher priority to highly connected nodes.

---

[3] Since a web graph is directed, there are two kinds of connectivity for any node: inlink connectivity (coming to the node) and outlink connectivity (outgoing from the node).

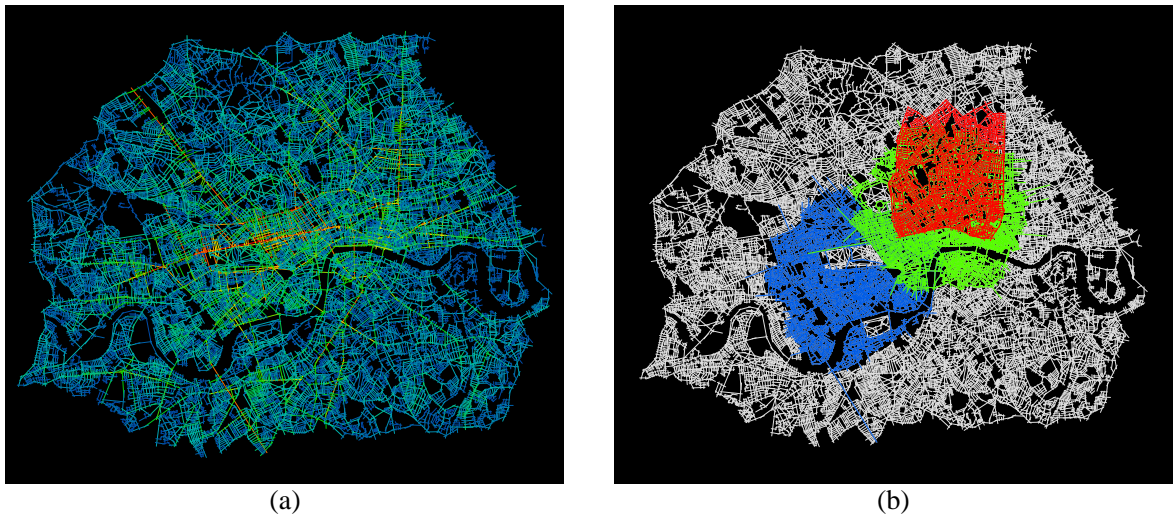

(a)            (b)

Figure 2: (color online) London axial map colored with local integration (a), and three test areas in central London (b): Barnsbury (red), Clerkenwell (green) and South Kensington (blue)

### 3.2 Small world and scale free structures

A first goal of the study is to assess the underlying topology of the London urban environment from a structural point of view. To this end, both path length and clustering coefficient were computed for the London topology, in comparison to its equivalent regular and random counterparts: (1) path length to assess how one node separates from all other nodes within the topology, and (2) clustering coefficient to indicate the level of clustering of the individual nodes of the topology. As we can see, the two metrics are initially defined as structural properties of individual nodes. However, they become a property of a topology by taking an average of the metrics for the individual nodes, which can be used to assess whether or not a real world network obeys the small world structure. Our computed result confirms the small world structure of the London axial map, i.e. on average the lines are shortly separated at a global level, and highly clustered at a local level. The illustrated small world property seems universal across the whole London region. Three test areas in central London exhibit the same kind of structure; although the average connectivity and path length differ from the entire topology to the three extracted ones (Table 1). The small world structure appears to suggest the kind of hidden geometry that Hillier et al. (1999) elaborated while arguing why space syntax works.

Table 1: Small world metrics for the London topology and its three subsets

| Area | N | M | L | C | Lreg | Lrand | Crand |
|---|---|---|---|---|---|---|---|
| Barnsbury | 1,915 | 4.7 | 8.4 | 0.18 | 638.3 | 4.9 | 0.0025 |
| Clerkenwell | 3,622 | 4.9 | 8.3 | 0.18 | 370.8 | 5.2 | 0.0014 |
| Kensington | 2,742 | 4.7 | 8.1 | 0.16 | 914.0 | 5.1 | 0.0017 |
| London | 17,321 | 4.2 | 15.9 | 0.17 | 5773.7 | 6.8 | 0.00024 |

N = the number of lines      Lreg = L for regular graph
M = the average connectivity      Lrand = L for random graph

L = path length  Crand = C for random graph
C = clustering coefficient

Another aspect of the so called hidden geometry, to my understanding, is that both connectivity and line length follow a power law distribution. A power law relationship between two variables *x* and *y* can be mathematically expressed as $y = kx^{-\alpha}$. A very typical power law is called Zipf's law, named after George Kingsley Zipf (1949) who popularized first the law for city size distributions, word frequencies, and income distributions. Zipf observed that the frequency of English words occurred unevenly in a text. A typical power law curve starts at its maximum frequency and decreases sharply to a certain value and then continues to decrease very slowly all the way to infinity. Eventually an L-shaped curve emerges for the power law. To examine whether or not a data set follows a power law, we can simply plot the data set in a log-log plot. If we take a logarithm for the above power law function, i.e. $Log(y) = -\alpha log(x) + log(k)$, clearly the L-shaped curve becomes a straight line on a log-log scale. Power law has been treated as a universal law, and has recently received a revival of research interests in an array of disciplines including for instance, physics, biology, sociology, and computer science (*cf.* Newman 2005). However, the ideal power law distribution is only applicable for a system with an infinite size in theory (Amaral et al. 2000). Our study illustrates that both connectivity and line length exhibit a sort of power law distributions. Figure 3 illustrates a series of log-log plots, whose x-axis represents connectivity or length, and y-axis cumulative probability. Apparently the log-log curves are pretty straight after a cutoff. Power law distribution is also called scale free property in the terms of Barabási and Albert (1999). This is not a particularly surprising finding, which has also been illustrated in previous studies (e.g. Carvalho and Penn 2004, Porta et al. 2006, Jiang 2007). However, none of the studies has particularly examined the distribution of the connectivity of axial lines.

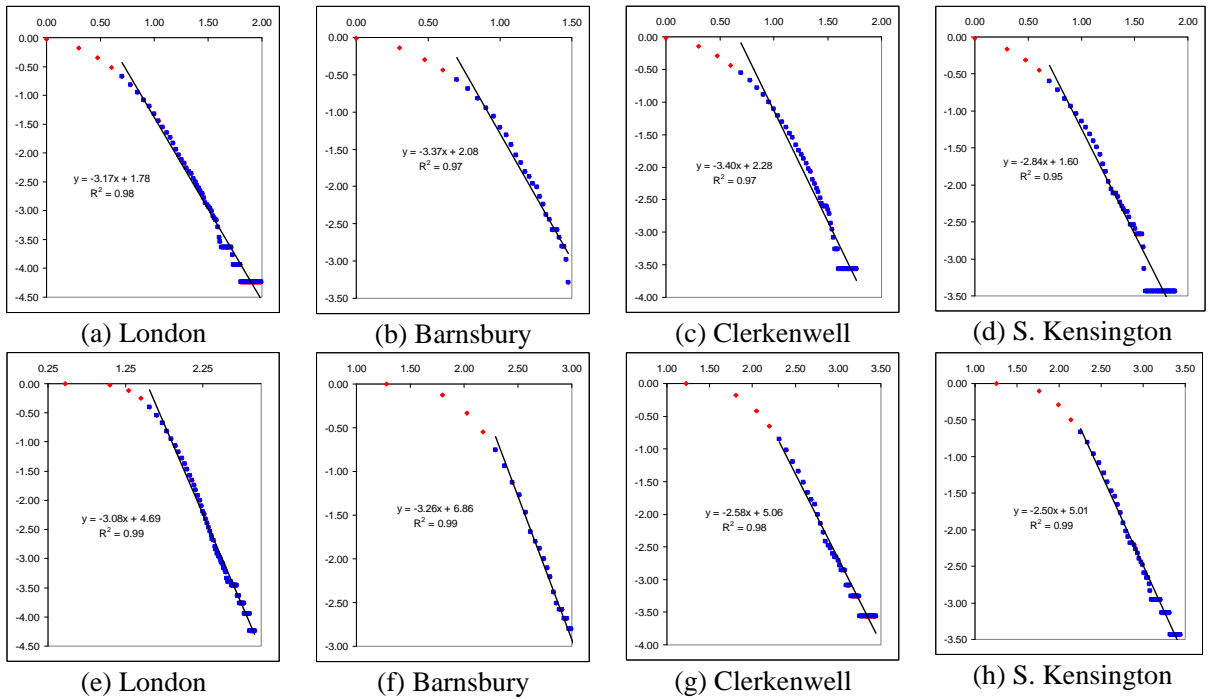

Figure 3: (color online) Power law distributions of connectivity (a-d) and length (e-h) of the axial lines of London and the three test areas

To this point, we have illustrated that the London topology exhibits both small world and scale free properties. In other words, the urban environment shows neither randomness nor order, but a hidden order, which is behind predictable human movement. This kind of structural properties resembles the semi-lattice that Alexander (1965) elaborated on, or what Hillier (1999) called hidden geometry. We will conjecture further that it is a major effect for predicting human movement by a random walker.

### 3.3 Correlation between human movement and ranking metrics
Having illustrated the hidden structure, in this section we will examine which ranking metrics have the best correlation to human movement. To this end, the dataset used for verification is observed pedestrian and vehicle flows for fifty minutes in almost all street segments around the four sites (Figure 4 and Figure 5). The number of gates set in the street segments varies from one site to another. The flows of individual gates are aggregated into their associated lines for correlation study. To our surprise, the weighted PageRank scores are significantly correlated to the pedestrian and vehicle flows, slightly better than local integration in the third decimal (Table 2). We have further noticed that the correlation between PageRank and movement gets better and better as the damping factor *d* increases. This observation is valid for both the standard PageRank and the weighted PageRank. Table 2 also indicates that the three space syntax metrics do correlate to human movement, and the difference of r-squares is very trivial. In contrast, the weighted PageRank with damping factor 1.0 is different from other PageRank scores with other values of the damping factor. Our study ignores other factors such as predominant land use, road width, and building height. However, if they are taken into account, there is no doubt that prediction would be significantly improved (Penn et al. 1998). The distributions of pedestrian and vehicle flows are positively skewed. Therefore, before checking correlation, the flow rates are taken by the fourth root to give them more normal distribution. This adjusting process is applied to any metric if its distribution is skewed.

Table 2: R-squares for the correlation between ranking metrics and human movement
(NOTE: # = number of gates, connect.= connectivity, local = local integration with two steps, global = global integration, PR85 = PageRank with damping factor equal to 0.85, PR100 = PageRank with damping factor equal to 1.0, WPR99 = weighted PageRank with damping factor equal to 0.99, WPR100 = weighted PageRank with damping factor equal to 1.0)

| Site name (mode, #) | Connect. | Local | Global | PR85 | PR100 | WPR99 | WPR100 |
| --- | --- | --- | --- | --- | --- | --- | --- |
| Barnsbury (ped. 47) | 0.69 | 0.70 | 0.45 | 0.64 | 0.68 | 0.68 | 0.70 |
| ------------ (veh. 37) | 0.57 | 0.64 | 0.55 | 0.57 | 0.57 | 0.61 | 0.62 |
| Clerkenwell (ped. 33) | 0.48 | 0.54 | 0.54 | 0.28 | 0.31 | 0.52 | 0.56 |
| ------------- (veh. 24) | 0.60 | 0.75 | 0.64 | 0.54 | 0.59 | 0.73 | 0.74 |
| Kensington (ped. 46) | 0.59 | 0.62 | 0.41 | 0.42 | 0.59 | 0.62 | 0.62 |
| ------------- (veh. 31) | 0.79 | 0.77 | 0.65 | 0.79 | 0.79 | 0.78 | 0.79 |
| Knightsbridge (ped. 67) | 0.32 | 0.38 | 0.49 | 0.29 | 0.32 | 0.36 | 0.38 |
| --------------- (veh. 47) | 0.49 | 0.60 | 0.48 | 0.45 | 0.49 | 0.6 | 0.62 |
| Mean | 0.566 | 0.625 | 0.526 | 0.498 | 0.543 | 0.613 | 0.629 |

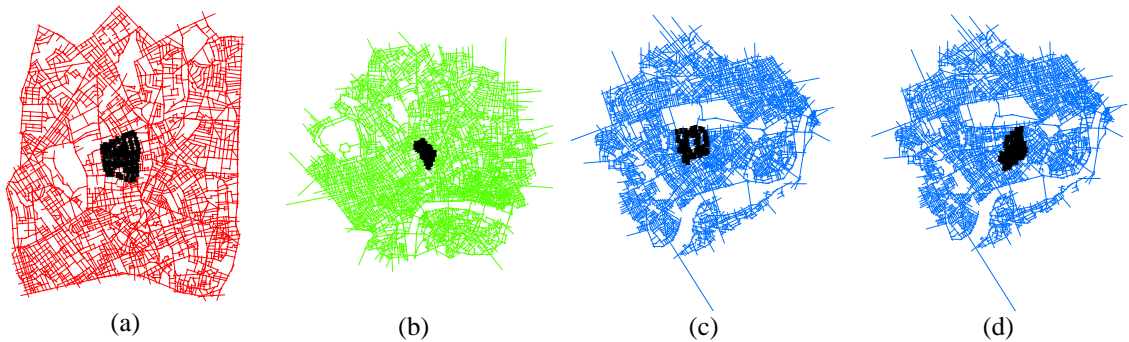

(a)    (b)    (c)    (d)

Figure 4: (color online) Four observed sites: Barnsbury (a), Clerkenwell (b), South Kensington (c) and Knightsbridge (d) embedded in the three central London areas

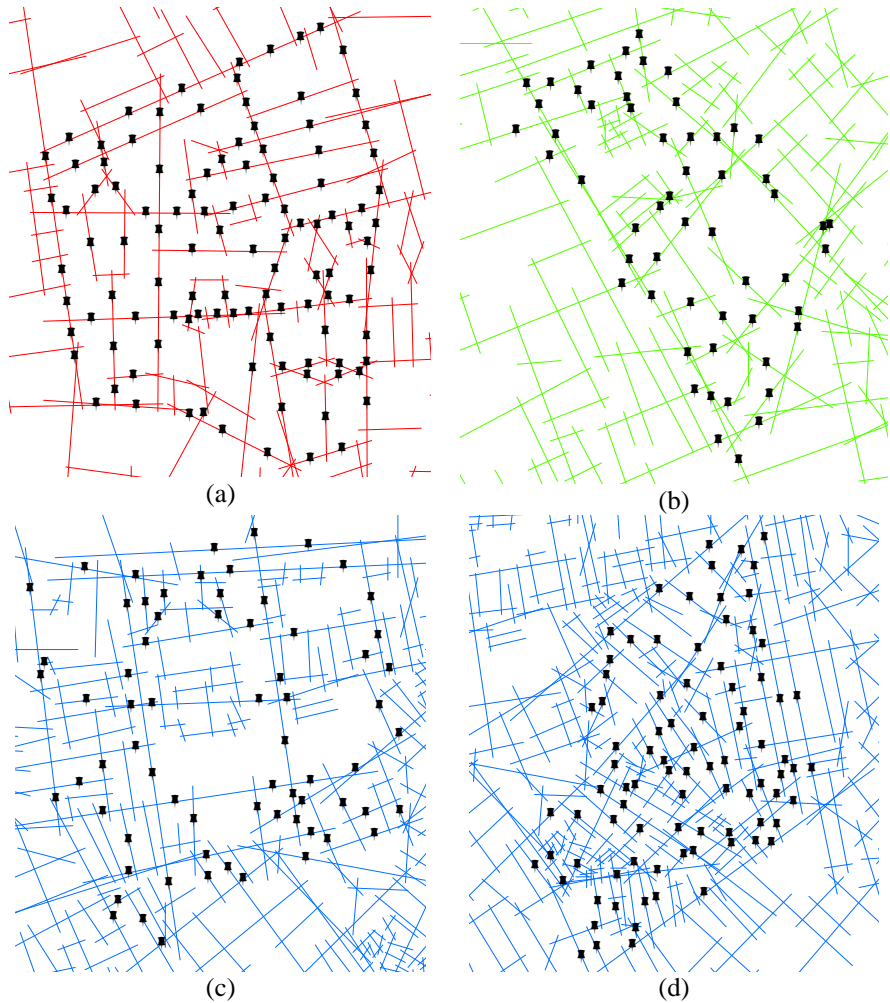

(a)    (b)
(c)    (d)

Figure 5: (color online) Enlarged view of the four observed sites: Barnsbury (a), Clerkenwell (b), South Kensington (c) and Knightsbridge (d)

**4. Justification of our findings**

The above experiments end up with two major findings: (1) the underlying space-space topology of an urban environment is far from random, but exhibits small world and scale free structures; (2) the weighted PageRank is significantly correlated to human movement. In this section, we will justify the findings from the point of view of the random walk (Woess 2000). In order to do so, it is important to note that the space-space topology differs from a web graph in two respects. Firstly, the topology is an undirected graph, while a web graph is directed. Secondly, because the topology is undirected, the topology contains no dangling nodes. The standard PageRank for a web graph is justified as a random surfer that has the following behavior:

*IF (a node has no successors) THEN*
*with probability 1 it jumps to a randomly chosen node*
*ELSE*
*with probability d it moves to one of its successors with a uniform probability, AND*
*with probability (1-d) it jumps to a randomly chosen node*
*END*

In words, the above behavior can be expressed in alternative ways. If the random surfer gets stuck, he will jump out to a randomly chosen node; otherwise, with probability *d* he moves to one of its successors with a uniform probability and with probability *(1-d)* he jumps to a randomly chosen node. Or simply stated, the behavior says that the random surfer jumps to a randomly chosen node when he gets stuck and tired; otherwise he moves to one of its successors with a uniform probability.

Because of the aforementioned differences between the space-space topology and a web graph, we justify our findings by a random walker. The random walker never gets stuck (as there are no dangling nodes in the space-space topology) or tired (as *d* is set to *1.0*), and always (i.e., with probability *1*) moves to one of its successors with non-uniform probability, i.e., well connected streets are more favorable. Thus the above weighted PageRank for predicting human movement in an urban environment can be simplified as follows:

$$PR(i) = \sum_{j \in ON(i)} PR(j)W(j) \qquad [9]$$

where W(j) keeps the same definition as in Equation [8].

Equation [9] provides a justification for our findings. It can be the justification as to why topological analysis be used to predict human movement within an urban environment. However, precaution should be taken seriously when we put the theory into practice. As we can see, over 60% of human movement can be predicted or explained purely from a topological point of view. In terms of statistics, the other 40% is not predictable, and it may relate to land use, building height and road width, etc. It is also important to point out that the 60% prediction is based on a corrected axial map which consists of the least number of longest visibility lines. However, we believe that the prediction can be significantly improved if we adopt a street-based topology.

**5. Conclusion and further work**
This paper has added new insights into ranking spaces or topological analysis of urban environments for predicting human movement. We examined the topological structure of London, and found that it exhibits small world and scale free structures. We further discovered that human movement is significantly correlated to the weighted PageRank scores, even slightly better than the local integration. Based on the PageRank random surfer model, we have justified our findings using a never-get-stuck-or-tired random walker who moves to its neighboring nodes (or successors) with non-uniform probability. This novel justification helps to answer the question as to why space syntax or topological analysis can be used to predict human movement in an urban environment. Clearly this justification is superior to the one suggested by Hillier (1999), i.e., people tend to take paths that minimize trip length or maximize trip efficiency. Our justification seems well supported both theoretically and empirically.

This study is based on the London datasets, which were collected using a very traditional data collection method, i.e., manually counting the number of vehicles or pedestrians through a gate. However, modern sensor technology provides far more advanced and accurate methods for data acquisition about human movement. We foresee that the deployment of the new technology will make new traffic datasets available for further verification. An observation derived from the study is that both local integration and weighted PageRank correlated significantly to human movement, but the two metrics did not correlate to each other. Also with respect to PageRank's damping factor, there is a sensitivity issue involved. All these issues and observations deserve further research in the future.

**Acknowledgement**
The study reported in the paper is financially supported by a Hong Kong Polytechnic University research grant. An earlier of this paper was presented at the XXIII International Cartographic Conference, 4-10 August 2007, Moscow, Russia. The author is grateful to Chengke Liu for his research assistance, and Amy Langville and David Gleich for their timely advice on efficient computation of PageRank. One of the referees provided constructive comments that better shaped the paper. However, any shortcomings are my sole responsibility.